\documentclass[pre,twocolumn,a4paper,nofootinbib,amssymb,amsmath,superscriptaddress]{revtex4}
\usepackage{graphicx,color}
\usepackage{epstopdf}
\usepackage{bm}
\usepackage{comment}
\usepackage{eucal}% altera il mathcal
\usepackage{mathrsfs}% per i caratteri col \mathscr

\begin{document}

\author{Damien Beecroft}
%\email{damien.beecroft@colorado.edu}
\affiliation{Department of Applied Mathematics, University of Washington, Washington 98195, USA}

\author{Juan G. Restrepo}
\email{juanga@colorado.edu}
\affiliation{Department of Applied Mathematics, University of Colorado at Boulder, Colorado 80309, USA}

\author{David Angulo-Garcia}
\email{dangulog@unicartagena.edu.co}
\affiliation{Universidad de Cartagena. Instituto de Matem\'{a}ticas Aplicadas. 
Grupo de Modelado Computacional - Din\'{a}mica y Complejidad de Sistemas. 
Carrera 6 \# 36 - 100, Cartagena de Indias, Bol\'{i}var - Colombia.}

\title{Greedy optimization for growing spatially embedded oscillatory networks}

\begin{abstract}
The coupling of some types of oscillators requires the mediation of a physical link between them,
rendering the distance between oscillators a critical factor to achieve synchronization.
In this paper we propose and explore a greedy algorithm to grow spatially embedded oscillator networks. 
The algorithm is constructed in such a way that nodes are sequentially added seeking to minimize the cost 
of the added links' length and optimize the linear stability of the growing network. We show that, for appropriate
parameters, the stability of the resulting network, measured in terms of the dynamics of small perturbations 
and the correlation length of the disturbances, can be significantly improved with a minimal added length 
cost. In addition, we analyze numerically the topological properties of the resulting networks and find that, 
while being more stable, their degree distribution is approximately exponential and independent of the 
algorithm parameters. Moreover, we find that other topological parameters related with network resilience 
and efficiency are also affected by the proposed algorithm. Finally, we extend our findings to more general classes
of networks with different sources of heterogeneity. Our results are a first step in the development 
of algorithms for the directed growth of oscillatory networks with desirable stability, dynamical and 
topological properties.
\end{abstract}

\maketitle

\section{Introduction}

The dynamics of large networks of coupled oscillators is of interest in many applications, including power grid systems 
\cite{witthaut2012braess,filatrella2008analysis,dorfler2013synchronization}, circadian rhythms \cite{lu2016resynchronization}, 
oscillatory brain rhythms \cite{kitzbichler2009broadband,breakspear2010generative}, and pedestrian synchronization 
\cite{strogatz2005crowd}. Finding characteristics of network structure that promote synchronization has been a subject of 
much research, and various techniques have been proposed to optimize the synchronization of oscillators coupled on a 
network \cite{skardal2014optimal, skardal2015control, skardal2016optimal,li2017optimizing, al2019optimization}. 
While coupled oscillator networks can often be analyzed by ignoring their spatial component, there are important cases where 
these networks are spatially embedded, including power grid systems \cite{witthaut2012braess,filatrella2008analysis,dorfler2013synchronization}, 
inner ear hair cells \cite{levy2016high,faber2021synchronization}, cortical circuits \cite{breakspear2010generative}, and electromechanical oscillators \cite{dou2018emergence}. 
In these cases, one should consider also spatial constraints when optimizing the synchronization of the oscillators.

Here we consider the problem of optimizing the synchronization of a growing network of spatially embedded oscillators while 
also minimizing the cost of the added connections. An illustrative example for our problem is the growth of electrical power 
grids. It is desirable for power grids to remain in a strongly synchronized regime as new nodes are added, while at the 
same time there is pressure to minimize the cost of the added lines. The cost of these lines depends on the geographical 
location of the added node and existing nodes. In this context, previous works have considered the growth of power grids by designing the addition of new nodes to optimize 
properties of the resulting network such as redundancy \cite{schultz2014random}, robustness to removal of nodes and path 
length \cite{pagani2016grid}, and other topological features such as variability in betweenness centrality and clustering 
coefficient (for more details see \cite{cuadra2017optimizing}). However, the interplay between the minimization of line 
costs and the need to optimize the stability of the synchronized state has not been explored (Ref.~\cite{al2019optimization} 
optimizes {\it synchronizability}, which is a related but different quantity).

In this paper we consider a growing network of coupled oscillators where new nodes are characterized by a stochastic geographical 
location, and the  connections to existing nodes are chosen so as to maximize the synchronization properties of the network and 
minimize the cost of the added connections. In contrast to previous works that focus on optimizing topological properties of the 
growing networks \cite{schultz2014random, pagani2016grid,cuadra2017optimizing}, we propose a greedy algorithm that directly optimizes 
the synchronization properties. More precisely, our algorithm optimizes a combination of the cost of the connections, 
taken to be proportional to the total Euclidean length of the network links, and a measure determinant of linear stability (a similar 
combination has been proposed for the growth of the internet \cite{fabrikant2002heuristically}). Remarkably, we find that by using 
our algorithm the stability and synchronization of the grown oscillator networks can be significantly improved without an appreciable 
increase in line length. 

Our paper is organized as follows. In Sec.~\ref{methods} we present our growing oscillator network model and the optimization algorithm. 
In Sec.~\ref{sec:properties} we analyze the topological and dynamical properties of the oscillator networks obtained from the growing algorithm. 
Next, in Sec. \ref{sec:heterogeneity} we show that the algorithm can also be applied to networks with different levels of heterogeneity.
Finally, we discuss our results and present our conclusions in Sec.~\ref{sec:discussion}.

\section{Model and Methods}\label{methods}

The growing oscillator network model is specified by the dynamics of individual oscillators and by the node addition process. The oscillator model will be presented in Section~\ref{sec:powerGrid} and the node addition process in Section~\ref{algorithm}.

\subsection{\label{sec:powerGrid} Oscillator Model and Stability}

For the dynamics of individual oscillators we will use the Kuramoto model with inertia \cite{filatrella2008analysis}, a rich oscillator model which, under some 
approximations (see Appendix \ref{appendixb}) can be used to model the dynamics of power grid systems. While the growing network process will be discussed in Section~\ref{algorithm}, for now we assume that the network has a fixed number $N$ of oscillators, where each oscillator is characterized by a phase $\theta_i$, $i = 1,2,\dots, N$, an intrinsic frequency $\Omega_i$, and a damping constant $\alpha_i$. The phase of oscillator $i$ evolves according to
\begin{equation}
\label{eq:kuramoto}
\ddot{\theta}_{i}(t)=\Omega_{i}-\alpha_{i}\dot{\theta}_{i}(t)+\sum\limits _{j}^{N}{\mathcal{K}_{ij}\sin[\theta_{j}(t)-\theta_{i}(t)]},
\end{equation}
where $\mathcal{K}_{ij}$ represents the coupling strength from oscillator $j$ to oscillator $i$. For simplicity, we will assume that $\mathcal{K}_{ij} = K A_{ij}$, where $K$ is constant and $A_{ij}$ are the entries of an $N \times N$
unweighted, symmetric adjacency matrix $A$. However, later we will discuss the case of weighted coupling matrices. By moving to a comoving rotating frame, 
we can assume without loss of generality that the average frequency is zero, $\langle \Omega \rangle = 0$. The state of each node can be represented by its phase angle $\theta_i$ and its angular velocity $\omega_i = d\theta_i/dt$. 

Depending on parameters, system (\ref{eq:kuramoto}) admits incoherent, partially, and fully synchronized solutions, and additional dynamical features such as hysteresis \cite{tanaka1997self, olmi2014hysteretic}. We will assume here that synchronization is desirable, and focus on the stability of the fully synchronized solution. For the example of power grids, fully synchronization is necessary for proper operation of the grids \cite{witthaut2012braess,filatrella2008analysis,dorfler2013synchronization}.
The fully synchronized solution is given by the fixed point $\omega_i = 0$, $d\omega_i/dt = 0$, corresponding to the phases $\theta_i = \theta_i^*$ that satisfy the equation
\begin{equation}
\label{eq:fixedpoint}
0 =\Omega_{i} + K\sum\limits _{j}^{N}{A_{ij}\sin(\theta_{j}^*-\theta_{i}^*)}.
\end{equation}

For small angle differences, the equilibrium can be approximated by
\begin{equation}
\bm{\theta}^* \approx \frac{1}{K} \mathcal{L}^{\dagger} \bm{\Omega}\label{eq:smallangle}
\end{equation}
where $\mathcal{L}^{\dagger}$ is the pseudo-inverse of the Laplacian matrix $\mathcal{L} = \text{diag}( {\sum_{j=1}^n A_{ij}} ) - A$, $\bm{\theta}^* = [\theta^*_1,\theta^*_2,\dots,\theta^*_N]^T$, and $\bm{\Omega} = [\Omega_1,\Omega_2,\dots,\Omega_N]^T$. In the case of weighted networks the definition of the Laplacian can be straightforwardly extended by replacing $A$ with $\mathcal{K}$.

The stability of the synchronized solution $\dot \theta_i = 0$, $\theta_i = \theta_i^*$ is determined by 
linearization of Eq.~\eqref{eq:kuramoto}. It has been shown in \cite{dorfler2013synchronization} that for a large class of network topologies 
a stable synchronized state with cohesive phases $|\theta_i^* - \theta_j^*|\leq \gamma < \pi/2$
can be achieved when
\begin{equation}
\label{eq:dorfler_sync_condition}
\Delta \equiv \frac{1}{K}  \|B^T \mathcal{L}^{\dagger} \bm{\Omega} \|_{\infty} < \sin(\gamma),
\end{equation}
where $B$ is the directed incidence matrix. Note that, taking the limit $\gamma \to \pi/2$ and recalling the small 
phase difference approximation of the equilibrium in Eq. \eqref{eq:smallangle}, Equation~\eqref{eq:dorfler_sync_condition} reduces to 
\begin{equation}
\label{eq:dorfler_sync_condition_2}
\Delta \approx  \|B^T \bm{\theta}^* \|_{\infty} < 1,
\end{equation}
which can be interpreted as saying that, 
in order to achieve stable synchronization, it is sufficient that the worst (largest) difference between the steady phase of connected pairs in the network 
is lower than $1$ \cite{dorfler2013synchronization}. The variable $\Delta$ is then an easily calculated index of stability, with a lower $\Delta$ being an indication of a more \textit{linearly} 
stable network \cite{galindo2020decreased}. 

In the next Section we present a network growth model where, each time a node is added, a combination of line cost and $\Delta$ is minimized by using a greedy algorithm. The main motivation for this problem is the growth of the power grid under the addition of power generation units (see Appendix \ref{appendixb}), but our results could be relevant for other situations where the stability of growing oscillator networks needs to be maintained.

\subsection{Spatial network growing algorithm}\label{algorithm}

In this Section we present the model for spatial network growth. In this model, nodes are sequentially added to the network at locations chosen stochastically from a prescribed probability density function. It is assumed that the addition of a new node has a cost that is proportional to the Euclidean length of the links used to connect it to the network, and that it is desired to minimize the total length of the added links (the {\it line length})  while maintaining the overall stability of the network. When a new node is connected to the network, a natural choice is to connect it to the closest nodes so as to minimize the added line lengths. However, here we propose that by connecting the new node to other nearby nodes, one can improve the stability of the network without significantly increasing the total line length. In the context of power grid modelling, there have been models for growing power grids that optimize network metrics such as robustness to node removal, assortativity, path length, and others \cite{schultz2014random, pagani2016grid,cuadra2017optimizing,li2017optimizing}; however, our model 
specifically addresses the optimization of a quantity that directly influences dynamical stability. We propose 
the following recursive spatial network growth model:

\begin{figure*}
\includegraphics[width=1\linewidth]{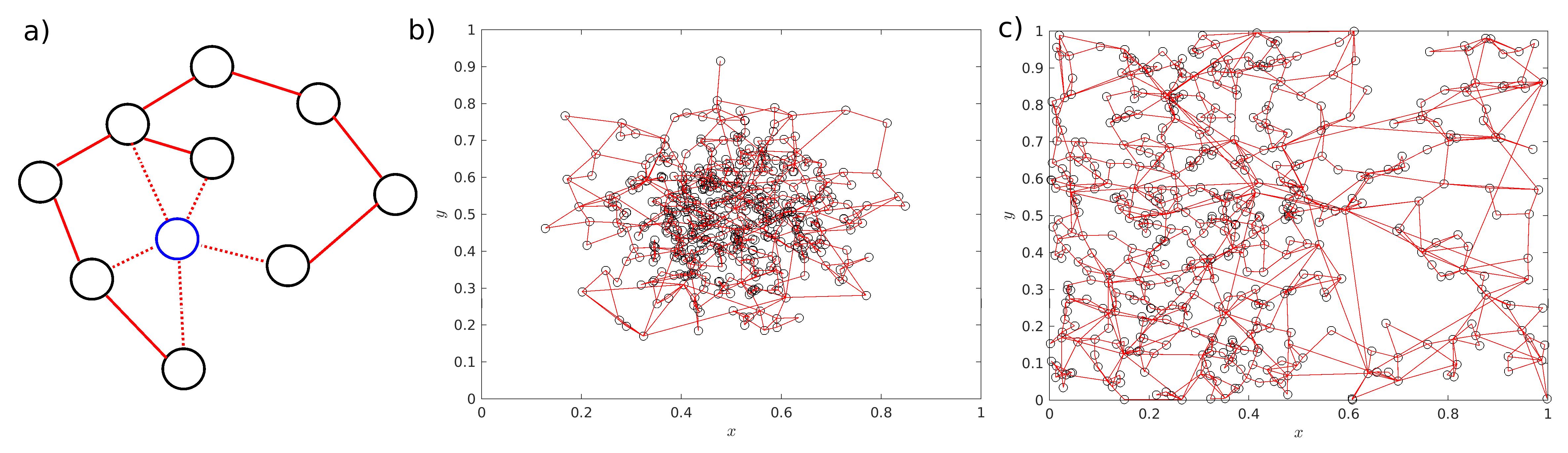}
\caption{a) A new node (blue) is placed at a position $(x_{t+1},y_{t+1})$ chosen randomly from the distribution $f(x,y)$. Potential connections (dashed lines) to the $q$ closest nodes are evaluated, and the $r$ connections with the lowest cost function $\mathcal{C}$ are established.
Networks constructed by following the growing algorithm with added node positions chosen b) with the $x$ and $y$ positions chosen independently from Gaussians centered at $0.5$ and with standard deviation $1/8$, and truncated so that the positions remain in $(0,1)$, and c) the $y$ position is chosen uniformly in $(0,1)$, and the $x$ coordinate is chosen from a piecewise constant distribution given by $8/5$ for $0<x<1/2$, $2/5$ for $1/2<x<1$, and $0$ otherwise. The other parameters of the algorithm are $s = 0.5$, $q = 5$, and $r =2$. The initial seed networks have $10$ nodes placed uniformly in the square $(0.4,0.6)\times(0.4,0.6)$ connected via their minimum spanning tree, and the final networks has $510$ nodes.}
\label{fig:newnode}
\end{figure*}

\begin{enumerate}

\item At time $t = 0$, the algorithm is initialized with a connected seed network of size $n_0$ spatially embedded in a simply connected region $M \subseteq \mathcal{R}^2$. Each node $i$ is characterized by coordinates $(x_i,y_i) \in M$ and an associated frequency $\Omega_i$ chosen in such a way that $\sum_{i=1}^{n_0} \Omega_i = 0$.

\item At time $t >0, t \in \mathbb{N}$, a new node is created with coordinates $(x_{n_0+t},y_{n_0+t})$ chosen randomly from a prescribed probability density $f(x,y)$ with support in $M$ and with associated frequency $\Omega_{n_0+t}$ chosen randomly from a probability distribution $g(\Omega)$. 

\item The frequencies are rebalanced so that the mean frequency remains zero. Motivated by power grid models where only generating nodes (those with $\Omega_j > 0$) can be adjusted, we modify only the positive frequencies as follows
\begin{align}
\Omega_i \to \left\{\begin{array}{cc}
\Omega_i,& \Omega_i \leq 0,\\
\Omega_i - \frac{\Omega_{t+1}}{N_+}\dots & \Omega_i > 0, i< n_0 + t,
\end{array}\right.
\end{align}
where $N_+$ is the number of previously existing nodes with positive frequency. 
We note, however, that a simple shift $$\Omega_i \to \Omega_i - \frac{1}{n_0 + t}\sum_{j=1}^{n_{0}+t}\Omega_j$$ produces similar results.
We also note that the zero average frequency condition can be relaxed as discussed in Sec. \ref{sec:heterogeneity}.

\item The newly added node establishes $r$ links to existing nodes, where the $r$ nodes are chosen among the closest $q$ nodes in such a way that the following cost function is minimized
\begin{equation}
\label{optimize}
\mathcal{C} = s \Delta + (1 - s) L,
\end{equation}
where $L$ is the total (Euclidean) line length after the new node is connected to the other $r$ nodes,
$\Delta$ is defined in Eq.~(\ref{eq:dorfler_sync_condition}), and $s \in [0, \, 1]$.

\item Steps 2-4 are repeated until a network of desired size $N$ is produced.
\end{enumerate}

The first term on the right hand side of the cost function [Eq. \eqref{optimize}] controls the degree of influence of the linear 
stability in the growing algorithm, while the second term controls the cost of establishing lines. A value of $s = 0$
seeks only to minimize the line cost and $s = 1$ seeks to enhance the linear stability of the resulting
network. Figure~\ref{fig:newnode}(a) illustrates the addition of a new node to the existing network (black circles with solid red lines). 
The new node (blue circle) is added at a random position, and potential links (dashed lines) to the $q = 5$ closest nodes are evaluated. 
The $r$ links that minimize $\mathcal{C}$ are established, and the procedure is then repeated with a new node.

Figs.~\ref{fig:newnode}(b-c) show two networks constructed by following the previous algorithm. In (b), the $x$ and $y$ positions 
are chosen independently from a Gaussian distribution centered at $0.5$, with standard deviation $1/8$, and truncated so that the positions $(x_i,y_i)$ 
remain within the region  $M = (0,1)\times (0,1)$. In (c), the $y$ position is chosen uniformly in $(0,1)$, and the $x$ coordinate is chosen from a piecewise 
constant distribution given by $8/5$ for $0<x<1/2$, $2/5$ for $1/2<x<1$, and $0$ otherwise. The other parameters are $s = 0.5$, $q = 5$, and $r =2$. The seed network consists of $10$ nodes placed uniformly in the square $(0.4,0.6)\times(0.4,0.6)$ connected via their minimum spanning tree. The frequency distribution $g(\Omega)$ here, and in the rest of the paper unless indicated, is uniform in $[-1,1]$.

\section{Dynamical and topological features of the growing networks}\label{sec:properties}

In this Section we show first how the algorithm can increase the stability of the grown networks with a negligible added cost. 
Then, we study additional dynamical characteristics of the grown networks such as linear stability and the correlation length 
of perturbations, and topological indicators such as degree distribution, clustering coefficient and betweenness centrality.

\subsection{Reduction of $\Delta$ with negligible cost}

\begin{figure}[t]
\includegraphics[width=\linewidth]{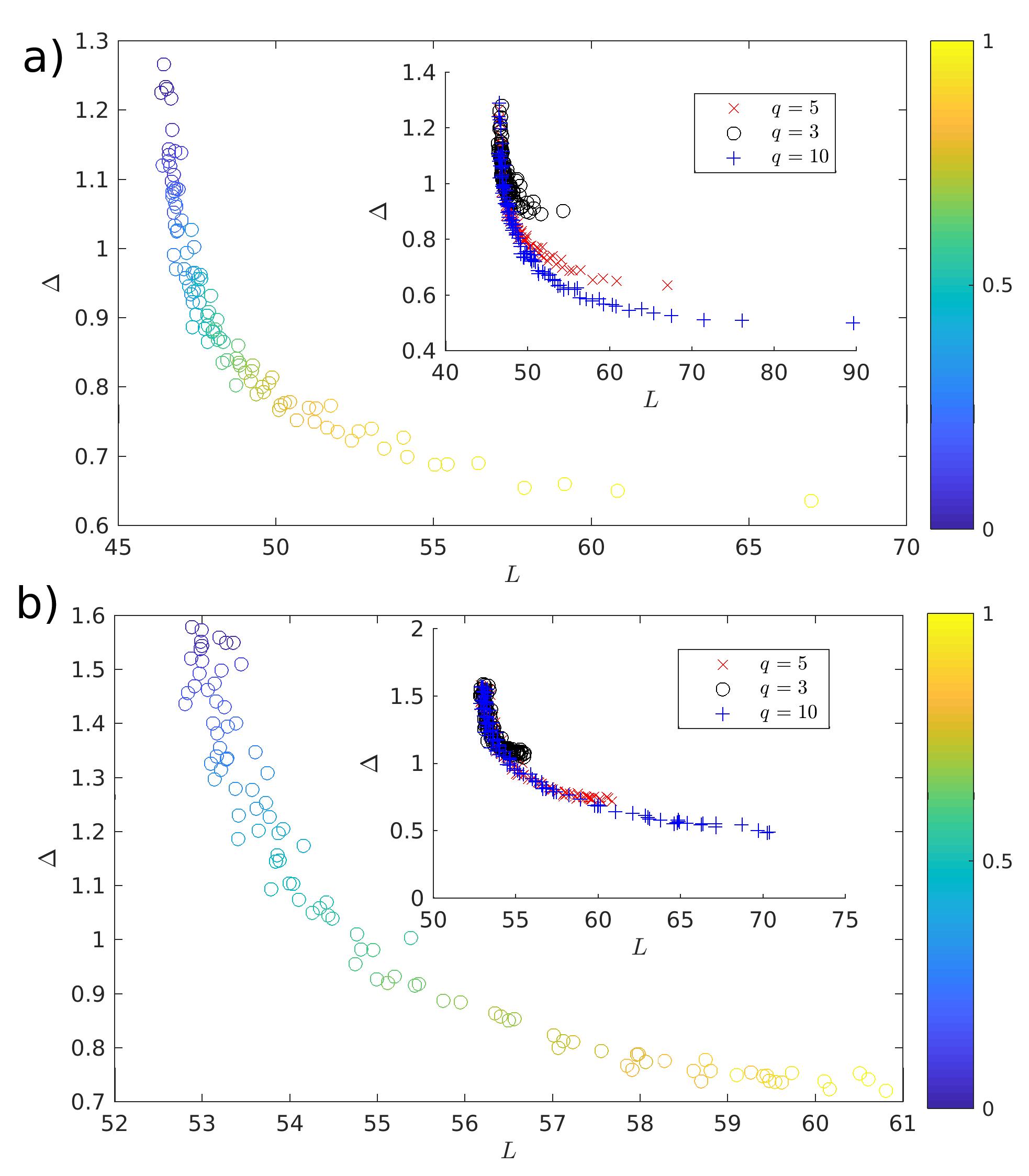}
\caption{a) Stability parameter $\Delta$ versus total line length $L$ averaged over $100$ realizations for $q = 5$, $r = 2$. The parameter $s$ is varied from $0$ to $1$ as indicated in the color bar. The inset shows the same results for $q = 3$ (black circles), $q = 5$ (blue x's), and $q = 10$ (red +'s). b) Results from the toy model
described in Appendix \ref{appendixa} showing similar qualitative results}
\label{fig:qs2}
\end{figure}

The basis of the growing algorithm is that, by allowing for connections to more distant nodes, the parameter $\Delta$ is reduced at the expense of increasing total line length $L$. Therefore, we expect that, as $s$ varies, $\Delta$ decreases as $L$ increases. This is verified in Figure \ref{fig:qs2}(a), where we plot $\Delta$ versus $L$ averaged over $100$ realizations as $s$ is varied from $0$ to $1$ (indicated by the color bar) for $q = 5$. The inset shows the same data for $q = 3$ (black circles), $q = 5$ (red x's), and $q = 10$ (blue +'s). While the plot confirms the above expectations, it also reveals the following behavior:
\begin{itemize}
\item Remarkably, for low values of $s$ there is a very sharp and significant decrease in $\Delta$ with an almost negligible increase in line length $L$.
\item For a {\it fixed} line length $L$, $\Delta$ decreases with increasing $q$. 
\end{itemize}
The first observation can be understood heuristically by considering the situation where a new node is added, and two potential connections are considered to nodes $i$ and $j$. If node $i$ is much more beneficial to minimize $\Delta$ than node $j$, but its distance to the new node is slightly larger than that of node $j$, a small but positive value of $s$ allows for the selection of node $i$ while only slightly increasing $L$. To understand the second observation, one can imagine all the possible ways in which a total line length $L$ is achieved. Since those with higher $q$ are obtained by allowing more potential connections, they allow for more chances to minimize $\Delta$ and should result therefore, on average, on a lower value of $\Delta$.

The above heuristic arguments are based solely on local considerations, and ignore the full complexity of how $\Delta$ depends on the network and the node parameters. To show that such local considerations can, indeed, result in the observed behavior, we considered a toy model where nodes are added sequentially, and the distances to and phase differences from potential connections are sampled from appropriate distributions (see Appendix \ref{appendixa} for details). This stochastic model reproduces qualitatively the numerical results as shown in Fig.~\ref{fig:qs2}(b).

In summary, although the growing algorithm is based on the competition between line length and stability, the results in Fig.~\ref{fig:qs2} show that one can improve stability without increasing the line length by (i) using small values of $s$, or (ii) increasing $q$ and adjusting $s$ appropriately.

\subsection{Reduction of critical coupling constant}

\begin{figure}[t]
\includegraphics[width=0.9\linewidth]{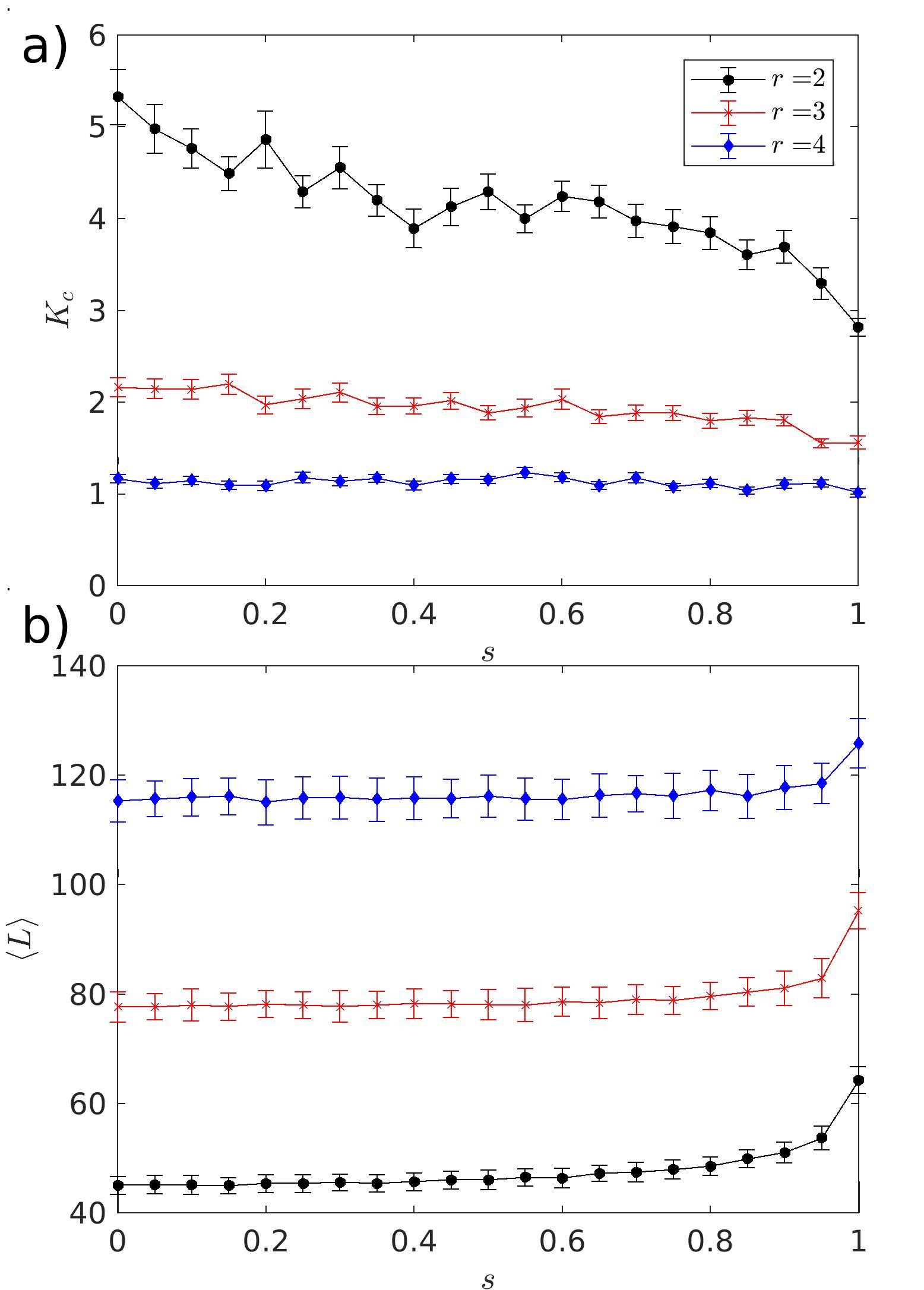}
\caption{a) The value of $K$ at which the synchronized fixed point loses stability, $K_c$, and b) the total line length $L$ as a function 
of $s$ for $r = 2$ (black circles), $r = 3$ (red x's), and $r = 4$ (blue diamonds). The symbols show an average over $100$ realizations and 
the bars represent one standard deviation.}
\label{fig:Kc}
\end{figure}

The growing algorithm is designed to minimize $\Delta$, which is a convenient indicator of linear stability. To study how the linear stability 
of the grown networks is actually improved, we perform the following numerical experiment: first we set $K$ at a value high enough such that the grown 
networks have linearly stable fixed points for all $s$ in $(0,1)$ (we used $K = 7$). For a given value of $s$, we grow a network of $N = 100$ nodes. 
Solving numerically Eq.~(\ref{eq:kuramoto}), the phases $\theta_i$ settle at their fixed point values $\theta_i^*$. Then, we adiabatically 
decrease $K$ until, at some value $K = K_c$, the system loses stability. The value of $K_c$ is averaged over $100$ realizations and the process 
is repeated for different values of $s$. The critical coupling strength $K_c$ is plotted versus $s$ in Fig.~\ref{fig:Kc}(a) for $q=5$ and $r  = 2$ 
(black circles), $3$ (red x's), and $4$ (blue diamonds). For $r = 2$ there is a significant reduction in the critical coupling as $s$ is increased, 
corresponding to a more linearly stable system. For $r = 3$ and $r = 4$, $K_c$ is smaller since there are more connections overall, but the reduction 
in $K_c$ as $s$ is increased is not as significant because the number of options when connecting a new node are reduced (e.g., there are $5$ options when making $r=4$ connections to $q = 5$ nodes, versus $10$ options when making $r=2$ connections to $q = 5$ nodes). Complementing the results shown in 
Fig.~\ref{fig:qs2}, we see that by increasing $s$ from $0$ to $0.85$ for $r = 2$, $K_c$ is decreased by approximately $40\%$ while the line length, 
shown in Fig.~\ref{fig:Kc}(b), increases only by about $10\%$.

\subsection{Linear stability}

Now we study the linear stability properties of the networks grown using our algorithm. While we have shown that higher values of $s$ reduce the critical coupling $K_c$ at which the fixed point $\theta^*$ becomes linearly unstable, here we show that, on the other hand, for high enough values of $K$ the linear stability properties of the grown networks are largely independent of $s$.
\begin{figure*}
\includegraphics[width=\linewidth]{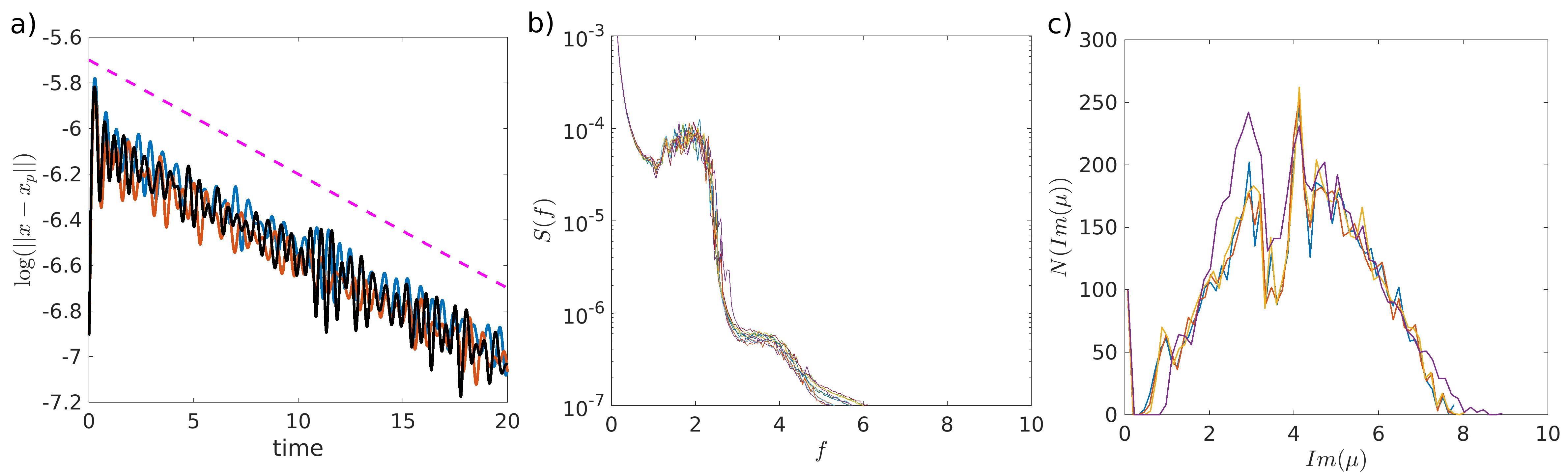}
\caption{a) Logarithm of the distance between the fixed point ${\bf x_p}$ and the perturbed trajectory ${\bf x}$  as a function of time. Blue, red 
and black curves correspond to three sample trajectories of networks generated with $s=0$, $s=0.5$ and $s=1$, respectively. The magenta line corresponds to a 
straight line with slope equals to $-\alpha/2$ showing that the decay of the perturbed trajectories towards the fixed point follows an exponential decay
with rate $-\alpha/2$. b) Frequency spectrum of the perturbations ${\bf x}_p$ for various values of $s$. c) Histogram of the imaginary part of the eigenvalues (\ref{eigenvalues}) of the state-dependent Laplacian  for various values of $s$.}
\label{fig:perturbation_dynamics}
\end{figure*}

\begin{figure*}
\includegraphics[width=\linewidth]{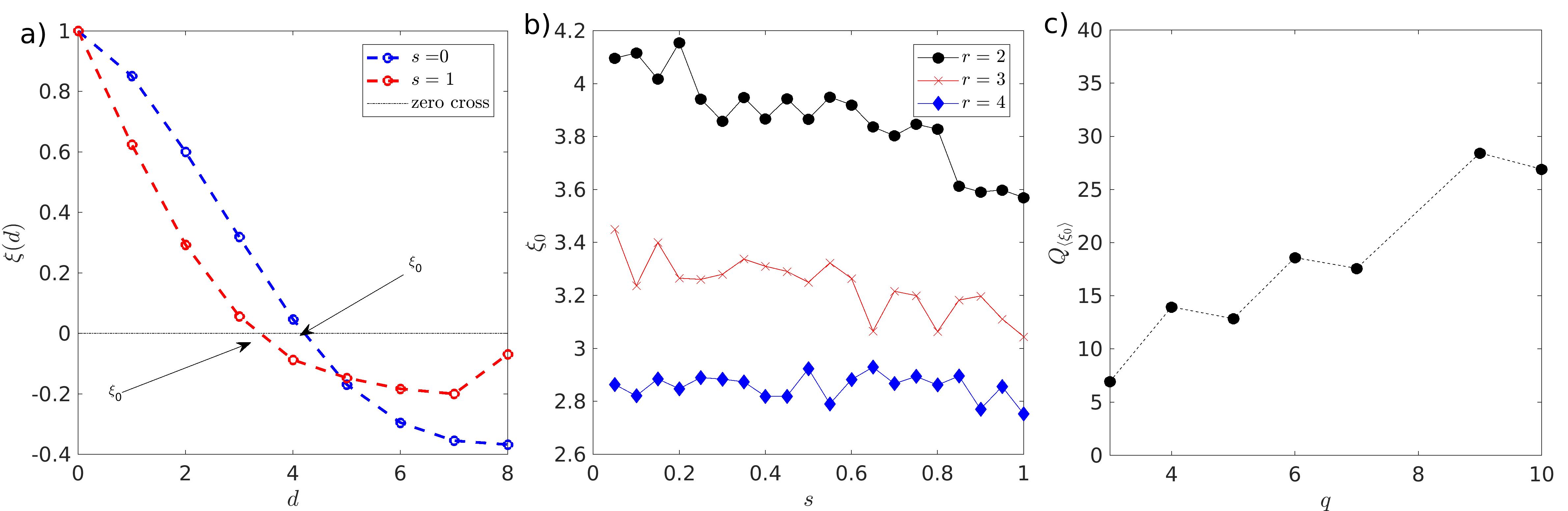}
\caption{a) Sample correlation length function for $s = 1$ (red) and $s = 0$ (blue) indicating the
first-zero crossing. b) First zero crossing of the correlation length function as a function of $s$ 
(symbols) for $K = 2$ at three different values of $r$ indicated in the legend of the figure. c) Relative change between $s=0$
and $s=1$ for different values of $q$ at a fixed value of $r$. We used $N_{seed} = 6$, $N = 94$, $q = 5$. We used $r=2$ in panels (a) and (c). The correlation length function is calculated averaging 20 iterations for each $s$.
\label{fig:correlation_length}}
\end{figure*}

The linearization around the equilibrium $\omega^*_i = 0$ and $\theta^*_i$ given by 
Eq. \eqref{eq:smallangle} of the system \eqref{eq:kuramoto} results in
\begin{eqnarray}
\label{eq:linearisation}
\delta \dot{\theta}_i &=& \delta \omega_i, \\
\delta \dot{\omega}_i &=& -\alpha \delta \omega_i - K \sum_{j=1}^N \mathcal{L}(\theta^*)_{ij}\delta \theta_j,
\end{eqnarray}
where  
$$
\mathcal{L}(\theta^*)_{ij} = \left\{\begin{array}{cc} 
 -A_{ij}\cos(\theta^*_j - \theta^*_i),&i \neq j,\\
- \sum_{k\neq i}^N L_{ik}, & i = j,
\end{array}
\right.
$$
are the entries of the so-called state-dependent Laplacian matrix $\mathcal{L}(\theta^*)$ \cite{li2017optimizing}. 
This shorthand notation allows us to write the Jacobian 
matrix of the system as
\begin{equation}
J = \begin{bmatrix}
\mathbf{0} & I \\[6pt]
- K \mathcal{L}(\theta^*) & - \alpha I 
\end{bmatrix},
\end{equation}
where $I$ is the $N\times N$ identity matrix. With this formulation, the eigenvalues of the Jacobian matrix can be
expressed as
\begin{equation}
\mu_i = -\frac{\alpha}{2} \pm \frac{1}{2}\sqrt{\alpha^2 - 4K\lambda_i(\mathcal{L}(\theta^*))}\label{eigenvalues},
\end{equation}
where $\lambda_i(\mathcal{L}(\theta^*))$ is the $i$th eigenvalue of $\mathcal{L}(\theta^*)$. Whether or not an eigenvalue $\mu_i$ has positive real part is determined by whether the eigenvalues $\lambda_i(\mathcal{L}(\theta^*))$ are all positive or not. When $A_{ij}|\theta_i^* - \theta_j^*| < \pi/2$ for all connected $i$, $j$, $\mathcal{L}(\theta^*)$ is diagonally dominated and positive semidefinite. In that case, and considering a low damping regime of the oscillators, all the eigenvalues $\mu_i$ have the same negative real part, $-\alpha/2$. 
The condition $A_{ij}|\theta_l^* - \theta_i^*| < \pi/2$ for all connected nodes $i$, $j$ is obtained when $\Delta \equiv \frac{1}{K}  \|B^T \mathcal{L}^{\dagger} \bm{\Omega} \|_{\infty} < 1$. Since $B$, $\mathcal{L}^{\dagger}$, and $\bm{\Omega}$ are independent of $K$, for large enough $K$ the fixed point $\theta^*$ is linearly stable, with Jacobian eigenvalues having identical and negative real part. 
To test this prediction, we generate networks at varying values of $s$ and fixed $K$. For each network, we perturb the nodal variables 
${\bf x} = ({\bf \theta}, {\bf \omega})$ from the synchronized fixed point ${\bf x}_p$ and  plot in Fig.~\ref{fig:perturbation_dynamics}(a) the logarithm 
of the euclidean distance between the perturbed trajectory and the fixed point, $\|{\bf x} - {\bf x}_p\|$, as a function of time for all networks.
From linearization one would expect that the distance evolves as  $\|{\bf x} - {\bf x}_p\| \propto \exp(\mu t)$, where $\mu$ is the leading
eigenvalue of the Jacobian. As seen in the Fig. \ref{fig:perturbation_dynamics}(a), the decay rate of the perturbations is independent of $s$ 
and approximately equal to $-\alpha/2$ (see magenta line with slope $-\alpha/2$).
This is not surprising as the real part of the eigenvalues is associated with the decay rate of the perturbations and this value is independent of $s$ as mentioned 
before. Interestingly, we also find that the frequency response of the perturbations, seen in the frequency spectrum [Fig.~\ref{fig:perturbation_dynamics}(b)] and 
the distribution of the imaginary part of the eigenvalues [Fig.~\ref{fig:perturbation_dynamics}(c)], are also largely independent of $s$. 
Thus, for large $K$, the linear response of the system does not depend on $s$. For moderate values of $K$, however, as shown in Fig.~\ref{fig:Kc}(a) 
and discussed earlier, the value of $s$ can be determinant for the linear stability of the fixed point.

\subsection{Correlation length function}

With the aim of further assessing the level of network resilience, we calculated 
the correlation length function of small (but finite) perturbations. Given a perturbation at a given
node, the correlation length function $\xi(d)$ is defined as the average correlation 
between the phase dynamics of every pair of nodes $(i,j)$ in the network at a 
topological distance $D_{ij}=d$. The topological distance for every pair of nodes in 
the network, in turn, is calculated as the length of the shortest path between them.
Altogether, the correlation length function reads as

\begin{equation}
\label{eq:corr_length}
\xi(d) = \frac{1}{N_d} \sum_{(i,j):D_{ij}=d} 
\left(\frac{\sum_t (\theta_i(t) - \bar{\theta}_i)(\theta_j(t) - \bar{\theta}_j)}
{\sqrt{\sum_t (\theta_i(t) - \bar{\theta}_i)^2(\theta_j(t) - \bar{\theta}_j)^2}}\right).
\end{equation}

Here $N_d$ is the number of pairs of nodes at a given distance $d$ and $\bar{\theta}_i$ 
is the time average of the phase $\theta_i(t)$. It is useful to 
calculate the first zero crossing of the correlation function ($\xi_0$) and use this as an 
indicator of how \textit{far} the effect of a perturbation propagates through the network. 

In Fig.~\ref{fig:correlation_length}(a) we report $\xi(d)$ for $s = 0$ (blue line)
and $s = 1$ (red line). For this test, we have assumed a connection 
strength $K = 2$ to guarantee consistent degrees of synchronization.
From this panel it is possible to see how networks generated via
a line-length optimization criteria ($s = 0$) have  a larger value of correlation length 
$\xi_0 \approx 4$, in contrast to networks generated following 
$\Delta$-minimization algorithm ($s = 1$), which gives $\xi_0 \approx 3$. This trend was
consistent across all the values of $s \in [0,\; 1]$ for $r=2$ and $r=3$, where a consistent
decrease of $\xi_0$ was found at increasing $s$ (see Fig. \ref{fig:correlation_length}(b)).
However, at $r=4$ there is virtually no difference between the correlation length at $s=0$ and $s=1$. 
To better understand the trend of $\xi_0$ at varying values of $q$, we introduced the
relative change of an indicator $x$ between its $s=0$ value and the $s=1$ value, namely:
\begin{equation}
\label{eq:Qx}
Q_{\langle x \rangle} = \frac{\langle x \rangle_{s=0} - \langle x \rangle_{s=1}}{\langle x\rangle_{s=0}}\times 100\%.
\end{equation}
In this equation, and in the following, $\langle x \rangle$ represents the average across realizations of $x$.
In the case of the correlation length $x \equiv \xi_0$ the result of this indicator is depicted in 
Fig.~\ref{fig:correlation_length}(c) where $Q_{\langle \xi_0 \rangle}$ increases from $5\%$ ($q=3$) to $\approx 25\%$ ($q=10$). 
This indicates that
the decreasing trend of $\xi_0$ with increasing $s$ is maintained by varying $q$. However, the changes are
relatively small.\\

In conclusion, decreased correlation length is a desired property of 
the network as it limits the extent of the effect of a perturbation at a given node. 
According to our analysis, this can be achieved with a $\Delta$-minimization scheme.

\subsection{Degree Distribution}

\begin{figure}
\includegraphics[width=0.9\linewidth]{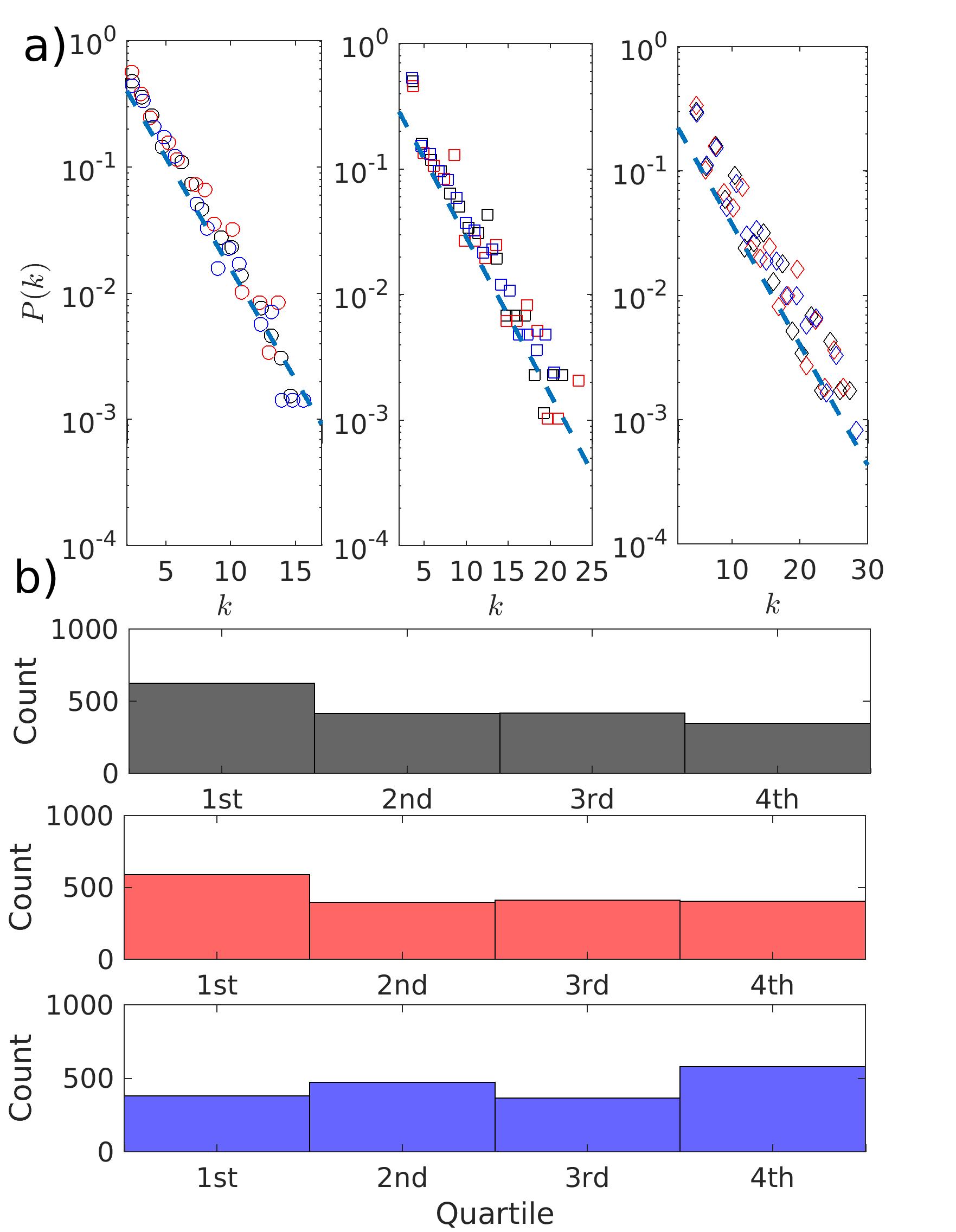}
\caption{a) Degree distribution of $N = 1000$ networks constructed using our model using $r = 2$ (left), $r=3$ (middle), and  $r=4$ (right) for $s = 0.0$ (black), $s = 0.5$ (red), and $s = 1.0$ (blue). b): quartile to which the degree of the nodes that new nodes connect to belongs to for $s = 0.0$ (black), $s = 0.5$ (red), and $s = 1.0$ (blue).}
\label{fig:degree_distribution}
\end{figure}

We proceeded to quantify some topological indicators to describe the resulting networks for different $s$. 
First we calculated the degree distribution, which fits an exponential function 
and is insensitive to the value of $s$. Figure \ref{fig:degree_distribution}(a) shows the degree distribution of networks constructed 
with $s = 0$ (black), $s = 0.5$ (red), and $s = 1$ (blue) with three different values of $r$. This type of distribution has been reported, for instance, in power grid connectivity in Ref.~\cite{deka2016analytical}. In the same reference, the authors considered a growth model in which nodes are placed spatially according 
to a two-dimensional Poisson point process with constant density and these are connected to the closest $r$ nodes 
[i.e., our model with $s = 0$ and constant $f(x,y)$]. 
Using a mean-field approach, the authors showed that the degree distribution of the resulting network has an exponential tail with 
exponent $\log[r/(1+r)]$. Remarkably, this theoretical estimate [dashed line in Fig.~\ref{fig:degree_distribution}(a)], valid in principle 
only for $s = 0$,  describes well the degree distributions obtained from our model with $s = 0.5$ and $s = 1$ as well. This can be understood 
by the empirical observation that when a node connects to the network, the choice of which $r$ nodes it connects to has very little correlation 
with the degree of these nodes as can be verified in Fig.~\ref{fig:degree_distribution}(b). For this figure, we perform one realization of network growth, storing at each 
growing step the quartile at which the degree of the $r$ connected nodes belong to. As seen in the Figure, the distribution of the quartiles 
is quite uniform, indicating the lack of correlation between the connected nodes and their degree.

Now we show that, using this assumption, the degree distribution is exponential with exponent $\log[r/(1+r)]$ even in the case that nodes 
are placed according to a non-uniform density $f(x,y)$. Let $n(x,y,k,t)$ be the density of nodes with degree $k$ at position $(x,y)$ 
at time $t$, and consider how the number of nodes of degree $k$ in a small region $S$ with area $\Delta A$ around $(x,y)$ is expected to change in one time step
\begin{align}
&n(x,y,k,t+1)\Delta A - n(x,y,k,t)\Delta A = \\
&n(x,y,k-1,t)\Delta A u \nonumber\\
&-n(x,y,k,t)\Delta A u,\nonumber
\end{align}
where
\begin{align}
u =  \frac{f(x,y)\Delta A r}{\sum_{k=r}^N n(x,y,k,t) \Delta A},
\end{align}
accounts for the probability that the added node is in the region $S$ [$f(x,y)\Delta A$], and the probability that it connects to a given node, obtained from the ratio of links established to the total number of nodes in $S$ [$r/\sum_{k=r}^N n(x,y,k,t) \Delta A$]. Simplifying, and approximating $n(x,y,k,t+1) - n(x,y,k,t) \approx dn(x,y,k,t)/dt$, we obtain the rate equation
\begin{align}
\frac{d n(x,y,k,t)}{dt} = \frac{f(x,y)r}{n(x,y,t)} [n(x,y,k-1,t)-n(x,y,k,t)],\label{eq:rate}
\end{align}
where $n(x,y,t) = \sum_{k=r}^N n(x,y,k,t)$. As $t\to \infty$, we look for a stationary solution of the form
\begin{align}
&n(x,y,k,t) = \bar n(x,y, k) t,\\
&n(x,y,t) = f(x,y) t.
\end{align}
Inserting this Ansatz in Eq.~(\ref{eq:rate}) and simplifying, we obtain
\begin{align}
&\bar n(x,y,k) = \frac{r}{1+r}\bar n(x,y,k-1),
\end{align}
so that the limiting distribution $\bar n$ is exponential
\begin{align}
&\bar n(x,y,k) = \bar n(x,y,r)e^{\ln\left(\frac{r}{1+r}\right)(k-r)}.
\end{align}

\subsection{Other topological indicators}

Although the degree distribution of the generated networks is insensitive to $s$, other topological properties are affected by the choice of $s$. We  computed other topological measures that characterize the generated networks, namely the average betweenness 
centrality of the network ($b$), the average clustering coefficient ($c$), and the characteristic path length ($l$), defined below:
\begin{eqnarray}
\label{eq:betweenness} 
b &=& \frac{1}{N}\sum_i \sum_{s,t \neq i} \frac{n_{st}(i)}{N_{st}},\\
\label{eq:clustering} 
c &=& \frac{1}{N}\sum_i \frac{ T_i }{ \mathcal{T} }, \\
\label{eq:path_length} 
l &=& \frac{1}{N(N-1)}\sum_{i,j} D_{ij}.
\end{eqnarray}
In Eqs.~\eqref{eq:betweenness}-\eqref{eq:path_length} $n_{st}$ is the number of shortest paths from nodes $s$ and $t$ that pass 
through $i$ and $N_{st}$ is the total number shortest paths from $s$ to $t$. $T_i$ is the number of triangles in which node $i$ 
is involved and $\mathcal{T}$ is the number of connected triplets in the network. Also, $D_{ij}$ is length
of the shortest path between the pair of nodes $(i,j)$.  At this level of description the differences between networks created at different weights $s$ start to emerge.\\

\begin{figure*}
\includegraphics[width=0.95\linewidth]{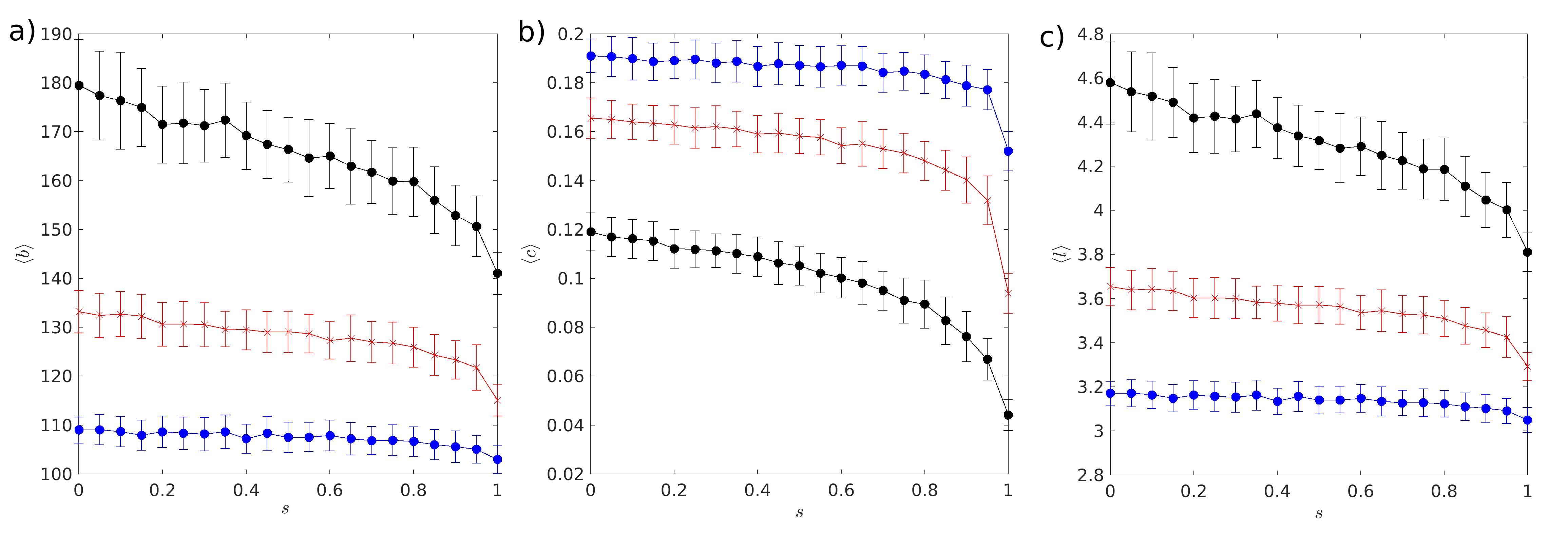}
\caption{Topological Indicators: Ensemble average of (a) betweenness centrality, (b) clustering coefficient, and (c) characteristic path length as a function of $s$. In all panels three values of $r$ were used, namely 
$r=2$ (black), $r=3$ (red) and $r=4$ (blue). For each value of $s$, 100 realizations 
of the algorithm were made with $N_{seed} = 6$ and $N = 100$. Error bars denote the standard deviation across realizations.}
\label{fig:topology_indicator}
\end{figure*}

\begin{figure}[b]
\includegraphics[width=0.95\linewidth]{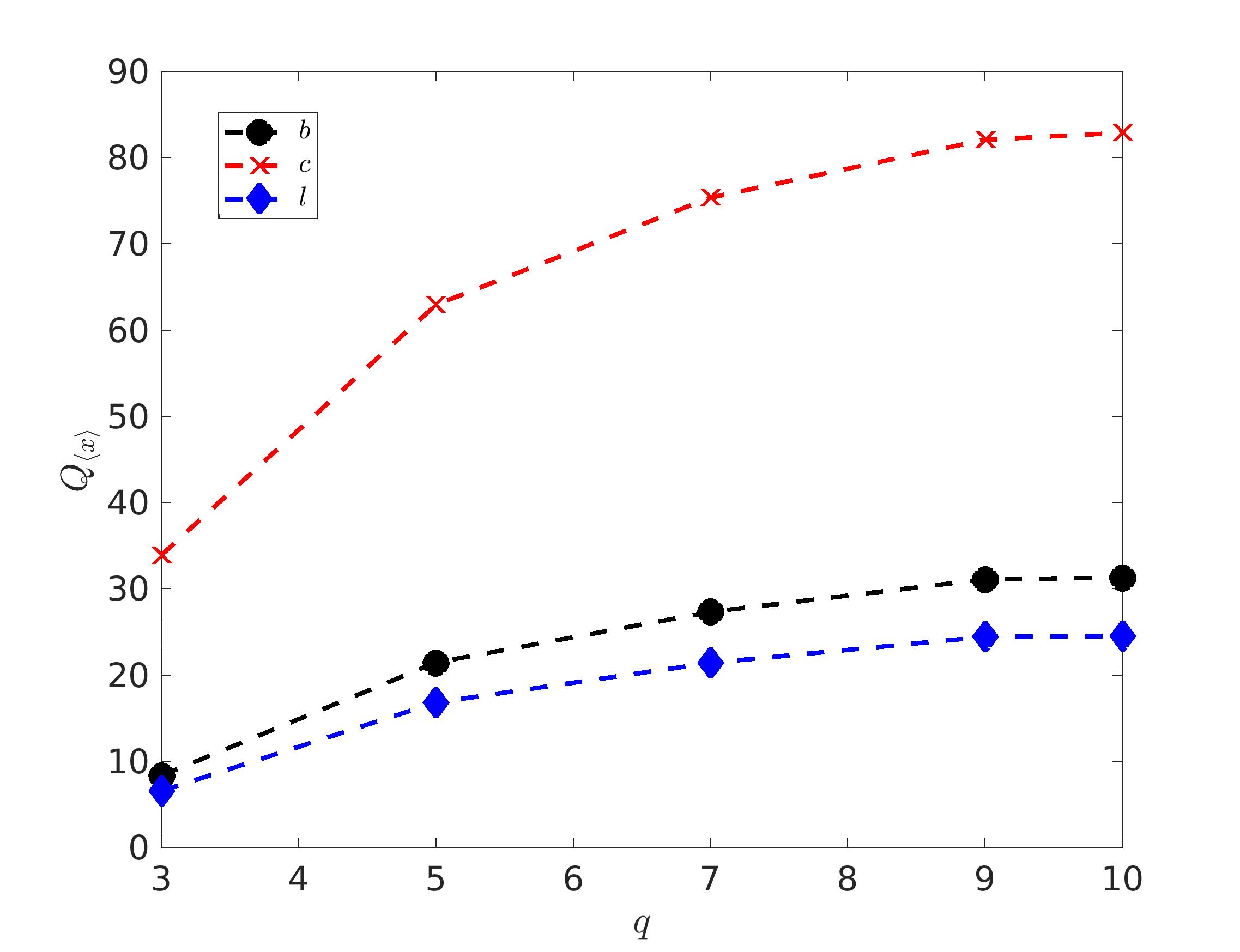}
\caption{Relative change between average topological indicators obtained at $s=0$ and $s=1$ expressed as a percentage. 
For this figure, a fixed value of $r=2$ was set, and then the relative difference between the topological indicator
obtained at $s=0$ and $s=1$ is expressed as a percentage for varying $q$.
Symbols as expressed in the legend. Parameters of network generation as in Fig. \ref{fig:topology_indicator}}
\label{fig:topology_indicator_varying_q}
\end{figure}

It has been proposed in \cite{cuadra2017optimizing,pagani2016grid} that resilient power grids 
are characterized by topologies with small values of average betweenness. 
The networks in our algorithm show a decreasing trend of $\langle b \rangle$ 
for increasing $s$ [see Fig.~\ref{fig:topology_indicator}(a)]
for the three considered values of $r$. This suggests that topological resilience is increased when seeking higher stability of the network. 
Conversely, the same authors showed that larger clustering coefficient and small characteristic path are indicators of efficient
power networks with reduced energy losses. From this perspective, the networks generated with our algorithm
tend to improve the characteristic path length with increasing $s$, while at the same time decreasing the clustering coefficient, as seen in Figs.~\ref{fig:topology_indicator}(b-c), indicating the need of a 
trade-off between resilience and effectiveness in our networks. It is worth noting that while the trends described above are maintained for all the 
values of $r$ studied, the relative differences between small and large $s$ are much more noticeable at low $r$. Of course the relative change 
between the topological indicators at small and large $s$ depends on the chosen value of $q$, namely the number of first neighbors that the 
greedy algorithm evaluates before choosing $r$ connections. To check this, we calculate the quantity $Q_{\langle x \rangle}$ defined in Eq.~\eqref{eq:Qx}, 
with $\langle x \rangle = \{\langle b \rangle,\; \langle c \rangle,\; \langle l \rangle \}$ which is depicted in Fig.~\ref{fig:topology_indicator_varying_q} 
by fixing $r=2$. In this Figure, it is possible to see that increasing the value of $q$, the relative change between $s=0$ and $s=1$ increases
for all the indicators, especially for the clustering coefficient where it changes from $40\%$ to $80\%$. Recall that, according to the definition in 
Eq.~\eqref{eq:Qx}, a positive value of $Q_{\langle x \rangle}$ is the result of a decreasing trend of the indicator at large $s$. From this, one can easily 
see that the clustering coefficient decreases more dramatically at large $q$. This is not surprising because the clustering coefficient reflects how 
well connected each node's neighbors are between them. Larger $q$ means that is it likely that neighbors are far apart, and therefore the chances 
that said neighbors are connected between them are lower. The results varying $q$ and $r$ seem to point out that considering more candidate nodes 
to connect to may have considerable effects on the efficiency of the network, as clustering is better achieved with local connections. This
preference towards local connectivity (decreased line length) should be however balanced with the dynamical features of the network encompassed by the
indicator $\Delta$.

\section{Effect of heterogeneity in the network}
\label{sec:heterogeneity}

\begin{figure*}
\includegraphics[width=0.95\linewidth]{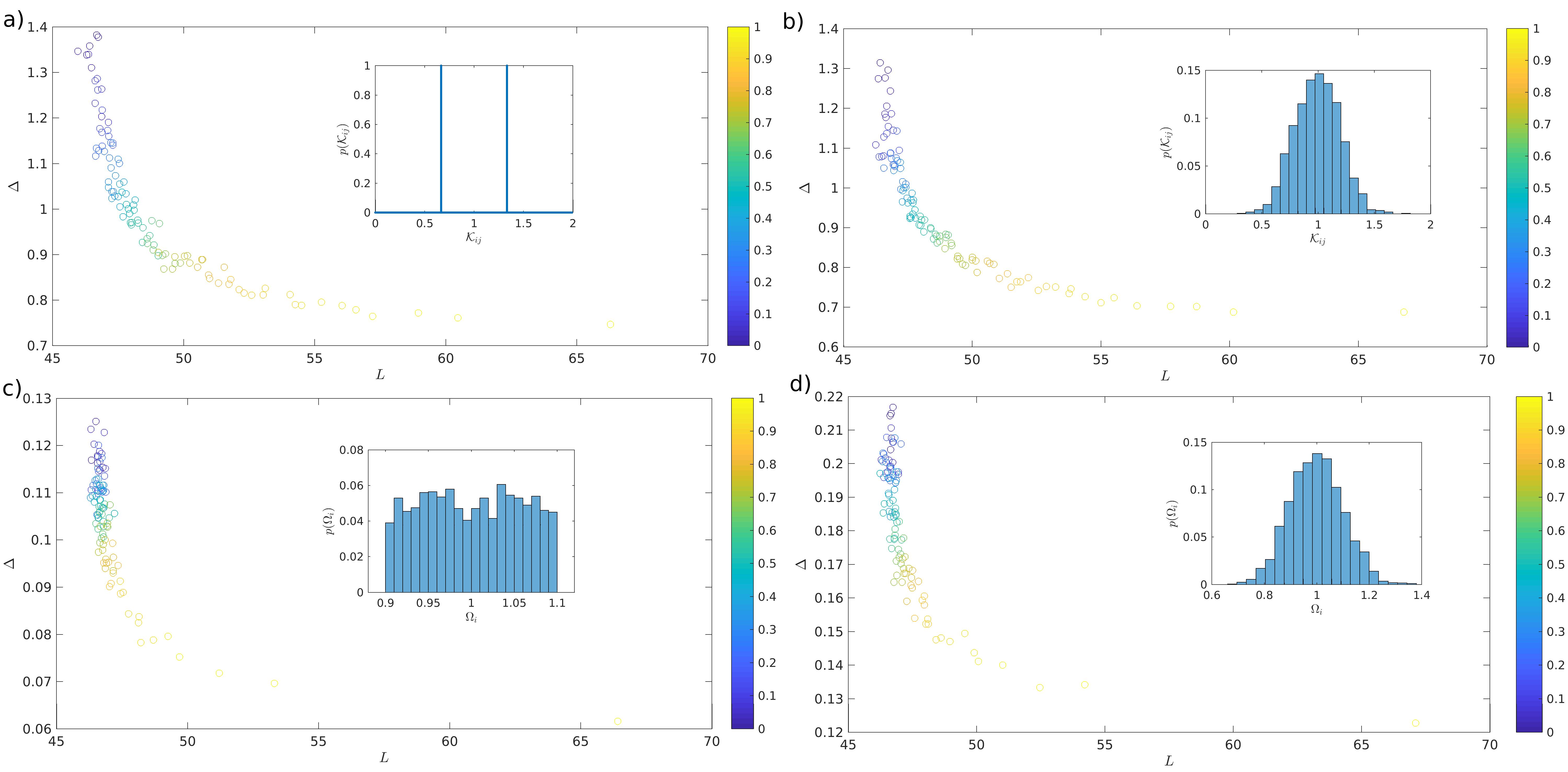}
\caption{Stability parameter $\Delta$ versus total line length $L$ averaged over $100$ realizations for $q = 5$, $r = 2$. The parameter $s$ is varied from $0$ to $1$ as indicated in the color bar. In a) for each newly created link its strength is chosen randomly from the discrete set $\mathcal{K} = \{2/3, 4/3\}$ with equal probability. In b) connection strength is chosen from a Gaussian distribution with average $\bar{\mathcal{K}} = 1$ and $\sigma(\mathcal{K}) = 0.2$. In c) the oscillator's frequency are chosen from a uniform distribution with support 
$\mathcal{U} \in [0.9 \; 1.1]$. In d) the oscillator's frequency is chosen for a Gaussian distribution with 
$\bar{\Omega} = 1$ and $\sigma(\Omega) = 0.1$.}
\label{fig:several_distributions_K}
\end{figure*}

Many real world networks have some degree of heterogeneity. For instance, in power grid networks the maximum capacity of the lines differs when passing from the high voltage 
transmission system to the power distribution system in populated centers. Also, in general oscillator networks, each node is usually described by a different natural oscillatory frequency. 
With this in mind, we studied the effect of heterogeneity when growing networks with our greedy algorithm. The results are summarized in Fig.~\ref{fig:several_distributions_K}. Panels a) and
b) show the averaged value of $\Delta$ as a function of $L$ at different values of $s$, considering that at each growing step the strengths of the connections $\mathcal{K}_{ij}$ are chosen according 
to a predefined distribution. In the case of panel a), the connection strength is chosen with equal probability from the discrete set $\mathcal{K}_{ij}=\{2/3,4/3\}$. This multimodal connection distribution is
inspired by the hierarchical nature of transmission lines in power transport systems. Similarly, panel b) was constructed choosing
at each iteration a connectivity strength drawn from a Gaussian distribution with mean value $\bar{\mathcal{K}} = 1$ and standard deviation $\sigma(\mathcal{K}) = 0.2$. For these two panels it is possible 
to observe that the general trend of the growing algorithm remains unchanged with respect to the main result discussed in Fig. \ref{fig:qs2}. Not only this, but also the range in which $\Delta$ varies
is quite similar in both cases and seem to be only driven by the average value $\bar{\mathcal{K}}$ which is identical in both distributions.

A second source of heterogeneity may come from the oscillator's natural frequency $\Omega_i$. In the case discussed in this work, $\Omega_i$ is drawn from a uniform distribution and 
imposing a frequency balance inspired by the behavior of power grids. With the aim of showing the generality of the approach proposed here, we also consider the case where $\Omega_i$ at each step
is drawn from different distributions without requiring zero average frequency condition. In panel c) we consider yet again uniformly distributed $\Omega_i$ with $\Omega \in [0.9 \; 1.1]$, 
i.e, disregarding step 3 of the algorithm. Similarly,
panel d) depicts the case in which $\Omega_i$ is drawn from a Gaussian distribution centered at $\bar{\Omega} = 1$ with standard deviation $\sigma(\Omega)=0.1$. As in the previous panels, the algorithm leads
to a similar trend, namely, there is an improvement of $\Delta$ with a negligible cost of $L$, however the actual values of $\Delta$ are now higher in the Gaussian distribution, despite the fact
that in both cases the average $\bar{\Omega} = 1$. This can be understood on the basis that $\Delta$ tends to be higher for networks with large variability of the intrinsic frequencies of the oscillators.
Although both distributions share the same mean, the variance of the Gaussian distribution is higher and therefore  the resulting networks are more heterogeneous. Despite these small differences, we
can conclude that the results presented in this work are general and can be applied to networks with different sources of heterogeneity.

\section{Concluding remarks}\label{sec:discussion}

In this paper we have proposed a greedy algorithm for the growth of oscillatory networks embedded in an 
Euclidean space, which uses the information of the added length and a readily available indicator of the 
linear stability of the resulting network. We have found that with a slight increase in the total added 
line we could obtain a significant improvement of the phase-cohesiveness of the network -a measure of 
the degree of stability of the synchronized state- and therefore network dynamical robustness.

Next, we studied the effect that the different growing protocols had on the linear stability properties of 
the system, measured by 
the critical coupling of the resulting networks and the eigenvalues of the Jacobian matrix. We showed that
the critical coupling can be substantially reduced when considering a growing protocol that seeks to minimize $\Delta$. 

Other approaches to reduce the critical coupling and improve phase-cohesiveness in Kuramoto complex networks 
have been proposed from an 
optimization perspective (see for instance \cite{skardal2014optimal,skardal2016optimal,fazlyab2017optimal}). These methods attempt to allocate the different
network properties (connectivity, frequency of the oscillators, weight of the edges) which optimizes a desired synchronization measure.
In contrast, our algorithm is based on purely local and step-wise measures based on real world constraints such as the spatial location
of the element of the network.

The analysis of the linear dynamical features of the system (dynamics around the equilibrium state), led to some surprising effects. For example, the dynamics 
of the network were virtually unchanged under different values  of $s$. Not only perturbations are damped at the same rate (an expected behavior from the 
spectrum of eigenvalues), but also the frequency component of the evolution of the perturbation remained unchanged with different $s$ (an effect that cannot be directly
concluded from the eigenvalue expression). Despite the evidence that $s$ does not affect the dynamics of small perturbations, it had a dramatic 
effect in decreasing the resulting critical coupling of the networks, a definitely desired attribute when stable synchronized dynamics is required, for instance in power grids.
Other approaches to the optimization of network stability properties have been studied before. For instance in \cite{li2017optimizing} 
the authors used variational equations to find connectivity values that enhanced network dynamics in terms of the real part of the eigenvalues, 
quantifying the rate at which the system is able to damp perturbations. It should be noticed that, in contrast with the cited reference, we used 
small values of $\alpha$ leading to complex eigenvalues with identical real part, and therefore a similar type of dynamics in terms of perturbation damping. 

The results on intermediate perturbation response showed that the extent to which perturbations are transferred to the network (correlation length) 
can be decreased by considering an optimization process taking into consideration the value of $\Delta$. 
Decreasing the correlation is a highly desired property which might mitigate cascading failures, a well known catastrophic effect in power grids 
\cite{duenas2009cascading,hines2009cascading}. Other approaches to assess network stability to finite perturbations in networks 
have been proposed in terms of basin stability in Power grids \cite{menck2013basin,menck2014dead}, Kirchhoff indices \cite{tyloo2018robustness}
and Finite Size Lyapunov Exponents \cite{cencini2013finite,angulo2014stable}. All these tools can be complementary and could lead to important new 
insights on the nonlinear nature of networks grown with our proposed algorithm.

We found that tuning the relative importance of the added length versus the dynamical stability 
of the network has little to no effect in the degree distribution of the resulting network. 
Indeed, the networks generated with the algorithm have all
an exponential degree distribution, as has been reported in the literature for several 
real-word power grids \cite{albert2004structural,deka2016analytical,sole2008robustness,crucitti2004topological,monfared2014topology,kim2017network}.
This is an important characteristic, as the resulting grid remains a single-scaled network, avoiding the
presence of hubs which heavily undermine network stability. 

We also analyzed the effect of the growing protocol on other topological features of the network which are also signatures
of network efficiency and resiliency that fall out of the two target variables minimized by the algorithm. In particular,
we saw that these two characteristics compete with each other when tuning the parameter $s$. This result advocates for 
more complex expressions in the cost function which may account for these features as performed in \cite{cuadra2017optimizing}.
However, it shall be noticed that our proposal contains the minimal ingredients that capture the two important elements to
account for in optimizing space embedded networks, namely topology and dynamics.

Finally, we analyzed the effect of heterogeneous parameters in the system. We showed that using heterogeneous coupling strengths
and natural frequencies leads to very similar results, indicating that the algorithm is robust and relies on a
strong theoretical support. This is not surprising as Eqs. \eqref{eq:fixedpoint} and \eqref{eq:dorfler_sync_condition} hold true
regardless of the underlying distributions of the connectivity matrix and the values of $\Omega$. We do not rule out the possibility 
that the inclusion of heterogeneity may have different effects on other measures studied in detail throughout the paper 
for homogeneously coupled networks. As a matter of fact, recent works have shown that heterogeneity in power grid networks
may affect nonlinear features of the network such as tripping times and basin stability \cite{montanari2020effects}.
This study is, however, out of the scope of this paper and could lead to interesting lines of research in the future.

\acknowledgments

Damien Beecroft was supported by the Undergraduate Research Opportunities Program at the University of Colorado at Boulder.
D.A-G would like to acknowledge the financial support by the \textit{Vicerrectoria de Investigaciones - Universidad de Cartagena} 
through Project No. 019-2021.

\appendix

\section{Relationship of Power Grid dynamics and the Second Order Kuramoto Model}\label{appendixb}

In this appendix we show that the second order Kuramoto model is equivalent under proper approximations to the dynamics of a Power Grid \cite{witthaut2012braess,filatrella2008analysis}. 
A power grid consists of $N$ rotating 
machines which either supply power to the grid (generators) or consume it (consumers). The dynamical state of the $i$-{th} machine can be quantified via its phase angle $\phi_i$ and 
its angular frequency $\dot{\phi}_i$. The machines in the grid operate at the same nominal value $\Omega$, and the phase deviation of the $i$-{th} machine with respect to the reference
angle $\tilde{\Omega} t$ is:

\begin{equation}
\label{eq:phase_diff}
    \theta_i = \phi_i - \tilde{\Omega} t
\end{equation}

Power balance requires that the power at the $i$-th node $P^m_i$ (generator or consumer) shall be equal to the sum of transmitted 
$P^t_i$, accumulated $P^a_i$ and dissipated $P^d_i$ components, i.e:

\begin{eqnarray}
\label{eq:power_balance}
P^m_i = P^t_i + P^a_i + P^d_i
\end{eqnarray}

Dissipated power is proportional to the square of the angular velocity $P^d_i = D_i \dot{\theta_i}^2$, where $D_i$ is a dissipation constant. Also, accumulated power is related
to the derivative of the kinetic energy of the machine via the relation $P^a_i = \frac{1}{2} I_i \frac{d (\dot{\phi_i})^2}{dt}$ with $I_i$ being the moment
of inertia. Finally, transmitted power between two connected machines $i$ and $j$ is proportional to the sine of the phase difference and the capacity of the transmission line connecting the
elements $\bar{P}_{ij}$, therefore $P^t_{i,j} = \bar{P}_{ij} \sin(\phi_i - \phi_j)=\bar{P}_{ij}\sin(\theta_i - \theta_j)$. 
Putting together these expressions in Eq. \eqref{eq:power_balance} we get:

\begin{eqnarray}
\label{eq:pow_bal_2}
P^m_i = D_i \dot{\phi_i}^2 + \frac{1}{2}I_i\frac{d(\dot \phi_i)^2}{dt} + \sum_j\bar{P}_{ij} \sin(\phi_i - \phi_j)
\end{eqnarray}

Recalling Eq. \eqref{eq:phase_diff} and using the fact that phase deviations are small compared with the grid frequency, that is, $\tilde{\Omega} \gg |\dot{\theta_i}|$, 
equation \eqref{eq:pow_bal_2} takes the form

\begin{eqnarray}
\nonumber I_i\tilde{\Omega}\ddot{\theta_i} = P_i^m - D_i\tilde{\Omega}^2 - 2D_i \tilde{\Omega}\dot{\theta_i} + \sum_j \bar{P}_{ij} \sin(\theta_j - \theta_i) \\
\end{eqnarray}

Redefining the parameters as:

\begin{eqnarray}
\Omega_i &=& \frac{P^m_i- D_i\tilde{\Omega}^2}{I_i\tilde{\Omega}}, \\
\alpha_i &=& \frac{2D_i}{I_i}, \\
\mathcal{K}_{ij} & = & \frac{\bar{P}_{ij}}{I_i\tilde{\Omega}},
\end{eqnarray}
leads to Eq.~\eqref{eq:kuramoto}. This model is formally known in engineering as the swing equation. Notice that, in defining the transmitted power we have considered 
lossless transmission. If resistance across transmission lines is included a slightly different second order Kuramoto model is obtained with a further phase shift in the sine term,
namely the transmitted power is proportional to $\sin(\theta_j - \theta_i - \gamma_{ij} )$, where $\gamma_{ij}$ is related with the angle between the real and imaginary part of the 
transmission line's impedance. This leads to the more general Kuramoto-Sakaguchi model. 

\section{Stochastic model for node addition}\label{appendixa}

In this Appendix we present a toy model that shows that the effects of the greedy optimization algorithm can be explained from local stochastic considerations. For simplicity, here we use a uniform Poisson process with density $\lambda$ for the placement of added nodes, i.e., we set 

\begin{equation}
    f(x,y) = \lambda = \frac{1}{\int_M dxdy}
\end{equation}, 

We start with $2$ nodes at time $t = 0$ and assume the initial angle difference $\delta_1 = \theta_1 - \theta_2$ is sampled  from a Gaussian distribution with mean $\sigma$. The value of $\sigma$ depends on the value of $K$ used: larger values of $K$ correspond to smaller $\sigma$. At time $t=0$ the line length is $L(0) = 0$, the stability parameter is $\Delta(0) = |\theta_1 - \theta_2|$, and the set of angle differences is $\mathcal{D}(0) = \{\delta_1\}$.

The model then proceeds recursively as follows:  at time $t =0,1,2 ,3,\dots$, when we already have a set of $n = 1+rt$ angle differences $\mathcal{D}(t) = \{\delta_1,\delta_2,\dots,\delta_{n}\}$, line length $L(t)$ and stability parameter $\Delta(t)$, we simulate the addition of a new node connected to $r$ existing nodes. We sample the distances $x_1,x_2,\dots,x_q$ from the new node to the $q$ closest nodes from the appropriate random variables that describe the 2-D point Poisson process.  For example, the distance $x_1$ to the closest node when $N$ nodes have been added has density 

\begin{equation}
f(x_1) = 2 N \pi \lambda x_1(1-\lambda \pi x_1^2)^{N-1}
\end{equation}. 

Similarly, we generate the potential angle differences $\tilde \delta_{n+1}, \tilde \delta_{n+2} \dots,\tilde \delta_{n+q}$ between the new node and the $q$ potential nodes from a Gaussian distribution with mean $\sigma$. Then we let $\bar \Delta_j = \max(|\tilde \delta_j|,\Delta(t))$, $L_j = L(t) + x_j$ and choose the $r$ nodes $i_1, i_2,\dots, i_r$ with the smallest cost function $s \bar \Delta_j + (1-s)L_j$. We then update the angle differences set to $\mathcal{D}(t+1) = \{\delta_1,\delta_2,\dots,\delta_{n}, \tilde \delta_{i_1},\tilde \delta_{i_2},\dots, \tilde \delta_{i_r}\}$, the stability parameter to $\Delta(t+1) = \max_{\delta \in \mathcal{D}}\{|\delta|\}$, and the line length to $L(t+1) = L(t) + x_{i_1}+\dots + x_{i_r}$. In Fig.~\ref{fig:qs2} we used $\lambda = 0.4$, $\sigma = 0.5$, and simulated the process until $t = 200$. Each point represents the average of $100$ realizations.

\end{document}